      [1999/12/01 v1.4c Il Nuovo Cimento]
\documentclass{cimento}

\usepackage{graphicx}  

\title{Gamma-ray binaries: microquasars and binary systems with pulsar}
\author{J.M.~Paredes\from{ins:1}}
\instlist{\inst{ins:1} Departament d'Astronomia i Meteorologia, Institut de Ci\`encies del Cosmos (ICC), Universitat de Barcelona (IEEC-UB), Mart\'{\i}' i Franqu\`es 1, 08028
Barcelona, Spain}
\PACSes{\PACSit{97.60.Gb, 97.60.Lf, 97.80.Jp}{Pulsars, Black holes, X-ray binaries}}
\begin{document}

\maketitle

\begin{abstract}
Several binary systems have been detected at High Energy (HE, $E > $100 MeV) and/or Very High Energy (VHE, $E >$ 100 GeV) gamma rays. Some
of them are X-ray binaries in which accretion feeds relativistic radio
jets and powers the non-thermal emission (i.e., microquasars), whereas in
others the power comes from the wind of a young pulsar instead of 
accretion. Although the power mechanism in these systems is different
(accretion vs pulsar wind), all of them are radio, X-ray and gamma-ray
emitters, and have a high-mass bright companion (O or B) star that is a
source of seed photons for IC scattering and target nuclei for hadronic
interactions. I review here some of the main observational results on the
non-thermal emission from X-ray binaries as well as some of the proposed
scenarios to explain the production of gamma rays.
\end{abstract}

\section{Introduction}
The type of binaries detected so far at VHE gamma-rays are high-mass X-ray binaries (HMXB) consisting of a donor star, an O or B star of mass in the range 8 -- 40 M$_{\odot}$, and a compact companion neutron star (NS) or black hole (BH). Until now, no low-mass X-ray binary (LMXB) has been detected at HE/VHE, suggesting that the bright primary star in HMXB plays an important role in the production of high-energy photons (the transient nature of LMXBs further complicates their possible detection at HE/VHE gamma rays). Some properties of these systems and an individual description of them have also been discussed in \cite{paredes08}. Binary systems of massive stars have also been proposed as sources of gamma rays \cite{benaglia03}, and there are hints at present that one of such systems, Eta Carinae, radiates in the GeV band \cite{tavani09a}.
At present, there are four X-ray binaries that have been detected at TeV energies. Three of them, PSR B1259$-$63 \cite{aharonian05b}, LS I +61 303 \cite{albert06} and LS 5039 \cite{aharonian05c} have been detected in several parts of their orbits and show a variable TeV emission. The other source, Cygnus X-1, showed evidence of TeV emission once during a flare \cite{albert07}. All these sources, as well as Cygnus X-3, have been detected above 100 MeV by {\it AGILE} \cite{sabatini10, tavani09b, tavani10} and/or {\itÊFermi}  \cite{abdo09c, abdo09a, abdo09b} satellites. 

Whereas the optical companion in all these sources is well known, the nature of the compact companion is unknown in three of them. While in the case of PSR B1259$-$63 it is clear that the compact object is a rapidly spinning non-accreting neutron star and the TeV emission is powered by rotational energy \cite{aharonian09, johnston92}, and in the case of Cygnus X-1 there is an accreting stellar-mass black hole, in LS 5039 and LS I +61 303 the uncertainties in the determination of the inclination of the systems \cite{casares05a, casares05b}  prevents to fix the mass of the compact object and, therefore, to know whether it is a BH or a NS.  In the case of Cygnus X-3, the strong absorption precludes the determination of the mass of the compact object. However, the accretion powered relativistic radio emitting jets observed in Cygnus X-3 allows to classify it as a microquasar together with Cygnus X-1.

Some properties of these systems are summarized in Table 1, and these sources are individually described below.

\section{Microquasars}
Cygnus~X-1 is the first binary system for which dynamic evidence for a BH was found \cite{gies86}. It is also the brightest
persistent HMXB in the Galaxy, radiating a maximum X-ray luminosity of a few times
$10^{37}$~erg~s$^{-1}$ in the 1--10 keV range. At radio wavelengths the source displays a $\sim$15~mJy flux density and a
flat spectrum, as expected for a relativistic compact jet (one-sided, with velocity $v>0.6c$) during the low/hard state
\cite{stirling01}.  
Arc-minute extended radio emission around Cygnus~X-1 was also found using the VLA \cite{marti96}. Its appearance was that of
an elliptical ring-like shell with Cygnus~X-1 offset from the center. Later, such structure
was recognised as a jet-blown ring around Cygnus~X-1 \cite{gallo05}. This ring could be the result of a strong shock that develops at the
location where the pressure exerted by the dark jet, detected only at milliarcsec scales, is balanced by the shocked ISM. The
observed radiation would be produced through thermal Bremsstrahlung by ionized gas behind the bow shock. 

MAGIC observed Cygnus~X-1, and evidence (4.1$\sigma$ post-trial significance) of TeV emission was found during a
short-lived flaring episode \cite{albert07}. This emission came from Cygnus X-1 and was unrelated to the ring-like structure. These TeV measurements were coincident with an intense state of hard X-ray
emission observed by {\it INTEGRAL}, although no obvious correlation between the X-ray and the TeV emission was found
\cite{malzac08}. The detection occurred slightly before the superior conjunction of the compact object, phase at which the
highest VHE $\gamma$-ray opacities are expected. After computing the absorbed luminosity that is caused by pair creation for
different emitter positions, it has been suggested  that the TeV emitter is located at the border of the binary system not to
violate the X-ray observational constraints \cite{bosch08, romero10}. A recent study of the opacity and acceleration models for the TeV
flare shows, under the assumption of negligible magnetic field, that an electromagnetic cascading model can explain
qualitatively the observed TeV spectrum, but not its exact shape \cite{zdziarski09}. 
Recently, {\it AGILE} detected significant transient gamma-ray  emission above 100 MeV from Cyg X-1 during a hard X-ray state \cite{sabatini10}.

 Cygnus~X-3 is a HMXB formed by a Wolf-Rayet
star and a compact object that is thought to be a neutron star for orbit
inclination angles above 60$^\circ$ or a black hole otherwise \cite{vilhu09}. 
Cyg~X-3 shows flaring radio levels of up to 20~Jy, and was first detected and closely
observed at this level in 1972, resulting in one of the best-known examples of
expanding synchrotron emitting sources. These outbursts can be modeled
successfully as coming from particle injection in twin jets \cite{marti92},
which have been subsequently imaged trough interferometric techniques
\cite{marti01,miller04}.

Long-term multiwavelength monitoring of Cyg~X-3 has revealed that strong
radio flares occur only when the source shows high soft X-ray flux and
a hard power-law tail. If the electrons responsible for the strong radio
outbursts and the hard X-ray tails are accelerated to high enough energies,
detectable emission in the $\gamma$-ray energy band is possible. 

The {\it AGILE}  \cite{tavani09b} and {\it Fermi} \cite{abdo09c} satellites detected transient gamma-ray emission above 100 MeV associated with Cygnus X-3. In particular, {\it Fermi}  detected its orbital period in gamma rays being the first time that a microquasar is unambiguously detected emitting high energy gamma-rays. These results show that Cygnus X-3 is a new HE gamma-ray source. At TeV energies, Cygnus X-3 has not been detected yet by the new generation of Cherenkov telescopes  \cite{aleksic10}.

\begin{table}
\begin{center}
\begin{narrowtabular}{0cm}{@{}l@{}l@{}l@{~}l@{}l@{~}l}
\hline
\\
{\bf Parameters} & {PSR~B1259$-$63} & {LS~I~+61~303} & {LS~5039} & {Cygnus~X-1} & {Cygnus~X-3}  \\[3pt]
\hline \\
System Type  & B2Ve+NS & B0Ve+NS$\?$  & O6.5V+BH$\?$ & O9.7Iab+BH & WN$_{\rm e}$+BH$\?$ \\[3pt]
Distance (kpc) & 1.5 & 2.0$\pm$0.2             & 2.5$\pm$0.5 & 2.2$\pm$0.2 & $\sim$7 \\[3pt]
Orbital Period (d)  & 1237 & 26.5  & 3.90603 & 5.6 & 0.2 \\[3pt]
$M_{\rm compact}$ (M$_{\odot}$)  & 1.4   & 1--4  & 1.4--5  & 20$\pm$5 & --\\[3pt]   
Eccentricity & 0.87 & 0.72      & 0.35$\pm$0.04 & $\sim$ 0 & $\sim$ 0 \\[3pt]
Inclination & 36 &  $30\pm 20$     & 20? & $33\pm 5$ & -- \\[3pt]
Periastron (AU) & 0.7 &  0.1  & 0.1 & 0.2 & --\\[3pt] 
Apastron (AU) & 10 &  0.7  & 0.2 & 0.2 & --\\[3pt] 

\hline
\\
{\bf Physical properties} & &   &  &  & \\[3pt]
\hline \\
Radio Structure  & Jet-like  & Jet-like   & Jet-like  & Jet
 + Ring & Jet   \\[3pt]
~~~~~~~~Size  (AU)  &120 &10--700 &10--$10^{3}$ &40 & $\sim 10^{4}$ \\[3pt]
Luminosity (erg s$^{-1}$)& & & & & \\
~~~~~$L_{\rm R(0.1-100~GHz)}$  & (0.02--0.3)$\times10^{31}$$^{\rm (*)}$ & 
(1--17)$\times10^{31} $ &$1\times10^{31}$   & $0.3\times10^{31}$ & $7\times10^{32}$ \\[3pt]
~~~~~$L_{\rm X(1-10~keV)}$  & (0.3--6)$\times10^{33}$  & (3--9)$\times10^{33}$  & (5--10)$\times10^{33}$  & 
$1\times10^{37}$ & $(3.9-7.9)\times10^{37}$ \\[3pt]
~~~~$L_{\rm VHE}$  & $2.3\times10^{33}$$^{\rm (a)}$  & $8\times10^{33}$$^{\rm (a)}$  & $7.8\times10^{33}$$^{\rm (b)}$ & $12\times10^{33}$$^{\rm (a)}$  & -- \\[3pt]
$\Gamma_{\rm VHE}$ & $2.7\pm0.2$ & $2.6\pm0.2$  & $2.06\pm0.05$ & $3.2\pm0.6$ & -- \\[3pt] 

\hline
\\
{\bf Periodicity} & &   &  &  & \\[3pt]
\hline \\
Radio &  48 ms and   &  26.496 d and  & persistent & 5.6 d & persistent and  \\[3pt]
 &   3.4 yr  &   4.6 yr &  &  & strong outbursts \\[3pt]
Infrared &-- & 27.0$\pm$0.3 d  & variable & 5.6 d & -- \\[3pt]
Optical & --&  26.4$\pm$0.1 d& -- & 5.6 d & --\\[3pt]
X-ray & variable& 26.7$\pm$0.2 d  & variable & 5.6 d & 0.2 d \\[3pt]
$>$ 100 MeV & variable& 26.6$\pm$0.5  & 3.903 & flare & 0.199655 d\\[3pt]
  &{\it AGILE} & {\it Fermi}& {\it Fermi}& {\it AGILE} &{\it AGILE}\&{\it Fermi} \\[3pt]
$>$ 100 GeV & variable & 26.8$\pm$0.2 d  & 3.9078 d& flare & -- \\[3pt]
  & HESS & MAGIC & HESS & MAGIC& \\[3pt]

\hline \\
\end{narrowtabular}
{\small $^{\rm (*)}$ Unpulsed radio emission} \\
{\small $^{\rm (a)}$ 0.2 $<$E$<$ 10 TeV} \\
{\small $^{\rm (b)}$ Time averaged luminosity.} \\
\caption{The five X-ray binaries that are MeV and/or TeV emitters. The reported orbital parameters of LS~5039 and LS~I~+61~303 are from \cite{casares05a, casares05b}. Slightly different parameters of these sources are reported in \cite{aragona09, sarty10}.}
\label{table}
\end{center}
\end{table}

\section{Binary systems with pulsars}
PSR~B1259$-$63/LS 2883 is a binary system containing a B2Ve main sequence donor, known as LS~2883, and a 47.7~ms radio pulsar orbiting
its companion every 3.4~years in a very eccentric orbit, with $e=0.87$ \cite{johnston94}. The radiation mechanisms and 
interaction geometry in this 
pulsar/Be star were studied in \cite{tavani97}. These authors concluded that the
star-pulsar wind interaction was the most feasible radiation powering mechanism. This was the first variable galactic source
of VHE gamma-rays discovered \cite{aharonian05b}. The TeV light curve shows significant variability and the observed
time-averaged energy spectrum can be fitted with a power law with a photon index $\Gamma_{\rm VHE}$ = $2.7\pm0.2_{\rm
stat}\pm0.2_{\rm sys}$. Different models have been recently proposed to explain these observations. In a hadronic
scenario, the emission and light curve at TeV, as well as in the radio/X-ray band, could be produced by collisions of high
energy protons accelerated by the pulsar wind and protons of the the circumstellar disk ($pp$), plus the emission from the
$pp$ secondary particles \cite{neronov07}. A leptonic scenario is presented in \cite{khangulyan07}, in which it is shown that
the X-ray and the TeV light curves can be explained by a synchrotron/IC scenario of non-thermal emission. {\it AGILE} has detected transient gamma-ray emission above 100 MeV from a source near the Galactic plane and positionally consistent with the binary pulsar PSR~B1259$-$63 \cite{tavani10}.

\begin{figure}[]
\includegraphics[scale=0.24]{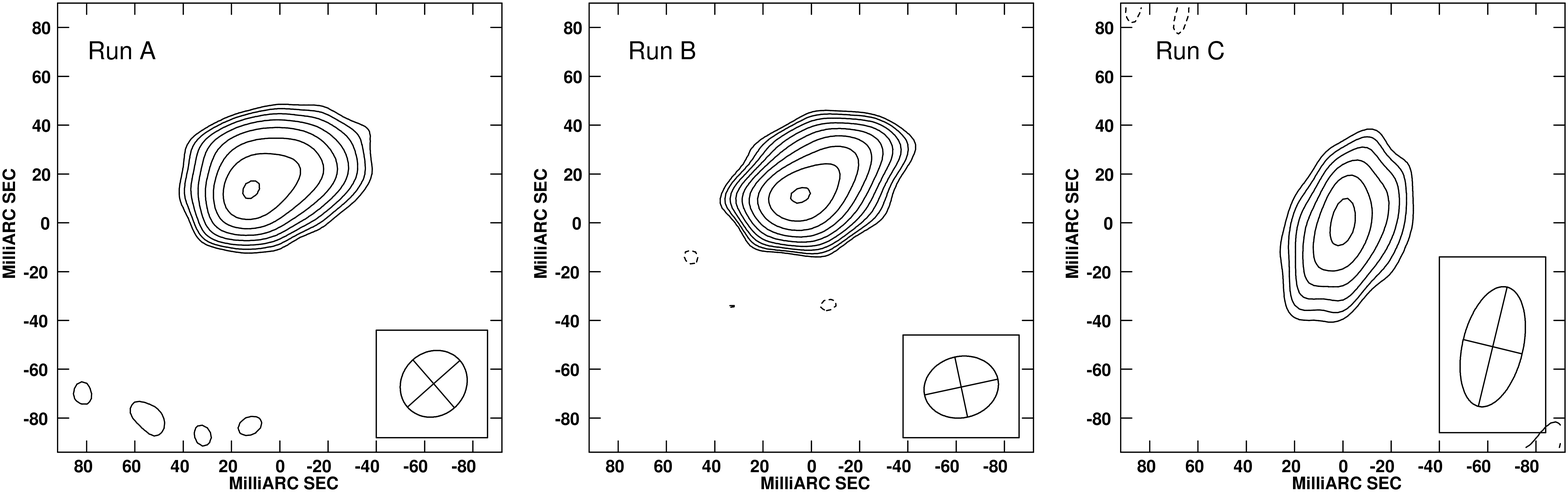}     
\caption{Australian Long Baseline Array (LBA) images of PSR B1259$-$63 at 2.3 GHz.  The synthesized beam is displayed in the rectangle on the bottom-right corner of each image. The size of the semi-major axis of the orbit of PSR B1259$-$63/LS 2883 is of the order of 3 mas. Figure from Mold\'on et al. (2010). }
\label{psr}
\end{figure}

Very recently,  high-resolution VLBI radio observations 
at three different orbital phases  of PSR~B1259$-$63 provided images (see Fig.~\ref{psr}) showing an extended and variable structure \cite{moldon10}.  Run A, B and C correspond to 1, 21 and 315 days after periastron passage respectively. The emission of run C is of a few mJy and is compatible with the flux density of the pulsar, giving the position of the pulsar. In run A and B we can see that there is an offset between the pulsar position and the peak emission of each image.  These results proof that non-accreting pulsars orbiting massive stars can produce variable extended radio emission at AU scales.

\section{Dubious cases}
The X-ray binaries LS~I~+61~303 and LS~5039 have been detected at HE and VHE. 
LS~5039 was detected by H.E.S.S.\cite{aharonian05c} and LS~I~+61~303 by MAGIC \cite{albert06}, being both systems periodic at TeV energies. {\it Fermi} has also detected emission modulated with the orbital period in both systems \cite{abdo09a, abdo09b}. The lack of strong evidence supporting the black hole or neutron star nature of the compact object in each of these systems does not allow their classification as microquasars or pulsar systems.

LS~I~+61~303 shows periodic non-thermal radio outbursts on average every $P_{\rm orb}$=26.496 days \cite{taylor82}.  In
\cite{massi04} it was reported the discovery of an extended jet-like and apparently precessing radio emitting structure with
angular extensions of 10--50~milliarcseconds. VLBA images obtained during a full orbital cycle show a rotating elongated
morphology \cite{dhawan06}, which may be consistent with a model based on the interaction between the relativistic wind of a
young non-accreting pulsar and the wind of the stellar companion (\cite{dubus06}; see nevertheless \cite{romero07} for a
critic review of this scenario).

The radio emission of LS~5039 is persistent, non-thermal and variable but no strong radio outbursts or periodic variability
have been detected so far \cite{ribo99, ribo02}. VLBA observations allowed the detection of an elongated radio structure,
interpreted as relativistic jets \cite{paredes00}. The discovery of this bipolar radio structure, and the fact that LS~5039
was the only source in the field of the error box of the EGRET source 3EG~J1824$-$1514 showing X-ray and radio emission, led
to propose their physical association \cite{paredes00}. High-resolution (VLBI) images of LS~5039 obtained five days apart have shown a changing morphology \cite{ribo08}. Precise phase-referenced VLBI observations covering
a whole orbital cycle are necessary for the detection of morphological and astrometric changes, which can be useful to
disentangle the nature of the compact source (see \cite{ribo08}). A theoretical discussion of the radio properties of LS~5039
can be found in \cite{bosch09}.

Some properties of LS~5039 and LS~I~+61~303 and an individual description of them can be found in \cite{paredes08}. Also, a
thorough  discussion of the different theoretical models for these systems is presented in \cite{bosch-khangulyan09}.  

\section{Summary}
Several binary systems have been detected at HE and/or VHE gamma rays. Two of them, Cygnus~X-1 and Cygnus~X-3, are accreting
X-ray binaries showing relativistic radio jets. A very different system detected at TeV is PSR~B1259$-$63, in which the power
comes from the pulsar wind and not from accretion. There are two other systems, LS~I~+61~303 and LS~5039, which have been
detected both at HE and VHE gamma-rays and where the nature of the compact object is not yet known. Although the powering mechanism of
these systems is different (accretion, pulsar), all of them are radio and X-ray emitters and have a high-mass bright
companion (O or B) star, which is a source of seed photons for IC scattering and target nuclei for hadronic interactions.

\acknowledgments
The author acknowledge support of the Spanish Ministerio de Educaci\'on y Ciencia
(MEC) under grant AYA2007-68034-C03-01 and FEDER funds.


\begin{thebibliography}{0}
\bibitem{paredes08}
\BY{Paredes, J.~M.}\TITLE{Proc. of Fourth Heidelberg International Symposium on High Energy Gamma-Ray Astronomy 2008 
(Heidelberg)} in \TITLE{2008 AIPC 1085, 157}

\bibitem{benaglia03}
\BY{Benaglia, P., \& Romero, G.E.} \IN{A\&A}{399}{2003}{1121}

\bibitem{tavani09a}
\BY{Tavani, M.,  et al.} \IN{ApJ}{698}{2009a}{L142}

\bibitem{aharonian05b}
 \BY{Aharonian, F. A., et al.}\IN{A\&A}{442}{2005b}{1} 
 
 \bibitem{albert06}
\BY{Albert, J., et al.}\IN{Science}{312}{2006}{1771} 

 \bibitem{aharonian05c}
\BY{Aharonian, F., et al.}\IN{Science}{309}{2005c}{746} 

 \bibitem{albert07}
\BY{Albert, J., et al.}\IN{ApJ}{665} {2007}{L51}

\bibitem{sabatini10}
\BY{Sabatini, S., et al.} \IN{ApJ}{712}{2010}{L10}

 \bibitem{tavani09b} 
\BY {Tavani, M., et al.} \IN{Nature}{462}{ 2009b}{620} 

\bibitem{tavani10}
\BY{Tavani, M., et al.}\IN{ATel}{2772}{2010}{}

 \bibitem{abdo09c} 
\BY {Abdo, A.~A., et al.} \IN{Science}{326}{2009c}{1512}      

\bibitem{abdo09a}
\BY{Abdo, A. A., et al.}\IN{ApJ}{706}{2009a}{706} 

\bibitem{abdo09b}
\BY{Abdo, A. A., et al.}\IN{ApJ}{701}{2009b}{L123} 

\bibitem{aharonian09}
\BY{Aharonian, F.A.,  et al.} \IN{A\&A}{507}{2009}{389}

\bibitem{johnston92}
\BY{Johnston, S.,  et al.} \IN{ApJ}{387}{1992}{L37}

\bibitem{casares05a}
\BY{Casares, J.,  et al.} \IN{MNRAS}{360}{2005a}{1105}

\bibitem{casares05b}
\BY{Casares, J.,  et al.} \IN{MNRAS}{364}{2005b}{899}

\bibitem{aragona09}
\BY{Aragona, C.,  et al.} \IN{ApJ}{698}{2009}{514}

\bibitem{sarty10}
\BY{Sarty, G.E.,  et al.} \IN{MNRAS}{tmp}{2010}{1789}




\bibitem{gies86}
\BY{Gies, D.R. \& Bolton, C.T.} \IN{ApJ}{304}{1986}{371}

\bibitem{stirling01}
\BY{Stirling, A.~M., et al.} \IN{MNRAS}{327}{2001}{1273} 

\bibitem{marti96}
\BY{Mart\'{\i}, J., et al.} \IN{A\&A}{306}{1996}{449}

\bibitem{gallo05}
\BY{Gallo, E., et al.} \IN{Nature}{436}{ 2005}{819} 

\bibitem{malzac08}
\BY{Malzac et al.} \IN{A\&A}{492}{2008}{527}

\bibitem{bosch08}
\BY{Bosch-Ramon, V., Khangulyan D. \& Aharonian F.~A.} \IN{A\&A}{489}{2008}{L21}

\bibitem{romero10}
\BY{Romero, G.E., Del Valle, M.V.  \& Orellana, M.} \IN{A\&A}{518}{2010}{12}

\bibitem{zdziarski09}
\BY{Zdziarski, A.A., Malzac, J., \& Bednarek, W.} \IN{MNRAS}{394}{2009}{L41}

\bibitem{vilhu09} 
 \BY{Vilhu, O., et al.} \IN{A\&A}{501}{2009}{679} 

 \bibitem{marti92} 
\BY {Mart\'i, J., Paredes, J.~M., \& Estalella, R.\ } \IN{A\&A}{258}{1992}{309} 
        
\bibitem{marti01} 
 \BY{Mart{\'{\i}}, J., Paredes, J.~M., \& Peracaula, M.} \IN{A\&A}{375}{ 2001}{476}

 \bibitem{miller04} 
\BY {Miller-Jones, J.~C.~A. et al.} \IN{ApJ}{600}{2004}{368}   
        
 \bibitem{aleksic10}
 \BY{Aleksi\'c, J., et al.}\IN{ApJ}{721}{2010}{843}
 
 \bibitem{johnston94}
 \BY{Johnston, S., et al.}\IN{MNRAS}{268}{1994}{430}
 
 \bibitem{tavani97}
 \BY{Tavani, M. \& Arons, J.}\IN{ApJ}{477}{1997}{439} 

\bibitem{neronov07}
 \BY{Neronov, N. \& Chernyakova, M.}\IN{Ap\&SS}{309}{2007}{253}

\bibitem{khangulyan07}
 \BY{Khangulyan, D., et al.}\IN{MNRAS}{380}{2007}{320}
 
\bibitem{moldon10}
\BY{Mold\'on, J., et al.}\IN{ApJ}{in press}{}{}
 
\bibitem{taylor82}
\BY{Taylor, A.~R, \& Gregory, P.~C.}\IN{ApJ}{255}{1982}{210}

\bibitem{massi04}
\BY{Massi, M., et al.}\IN{A\&A}{414}{2004}{L1} 

\bibitem{dhawan06}
\BY{Dhawan, V., et al. }\IN{
in Proc. of the VI Microquasar Workshop, Como}{}{2006}{}

\bibitem{dubus06}
\BY{Dubus, G.}\IN{A\&A}{456}{2006}{801} 

\bibitem{romero07}
\BY{Romero, G.~E. et al.}\IN{A\&A}{474}{2007}{15}

\bibitem{ribo99}
\BY{Rib\'o, M., et al.}\IN{A\&A}{347}{1999}{518} 

\bibitem{ribo02}
\BY{Rib\'o, M., et al.}\IN{A\&A}{384}{2002}{954} 

\bibitem{paredes00}
\BY{Paredes, J.~M., et al.}\IN{Science}{288}{2000}{2340} 

\bibitem{ribo08}
\BY{Rib\'o, M., et al.}\IN{A\&A}{481}{2008}{17}

\bibitem{bosch09}
\BY{Bosch-Ramon, V.}\IN{A\&A}{493}{2009}{829}

\bibitem{bosch-khangulyan09}
\BY{Bosch-Ramon, V. \& Khangulyan, D.}\IN{IJMPD}{18}{2009}{347}

        
\end{thebibliography}
\end{document}